\documentclass[a4paper,11pt]{article}
\usepackage[utf8x]{inputenc}

\usepackage{a4wide}
\usepackage{amssymb}
\usepackage{amsmath}
\usepackage{amsthm}
\usepackage[mathscr]{euscript}
\usepackage[normalem]{ulem}
\usepackage[colorlinks=true]{hyperref}
\usepackage[usenames,dvipsnames]{xcolor}
\usepackage{todonotes}
\usepackage{appendix}
\usetikzlibrary{arrows,angles}
\usetikzlibrary{matrix}
\usetikzlibrary{snakes}
\usetikzlibrary{arrows,calc,shapes,decorations.pathreplacing}
\usepackage{tikz-cd}

\numberwithin{equation}{section}

\theoremstyle{definition}

\newtheorem{Definition}{Definition}[section]

\newtheorem{Theorem}[Definition]{Theorem}
\newtheorem{Proposition}[Definition]{Proposition}
\newtheorem{Lemma}[Definition]{Lemma}
\newtheorem{Corollary}[Definition]{Corollary}
\newtheorem{Remark}[Definition]{Remark}
\newtheorem{Example}[Definition]{Example}
\newtheorem{Conjecture}[Definition]{Conjecture}

\newcommand{\R}{\mathbb R}

\usepackage[xcolor]{changebar}
\cbcolor{blue}
\newcommand{\Ric}{\mathrm{Ric}}

\usepackage[inline]{enumitem}
\newcommand{\enumlabelformat}{\roman}
\newcommand{\enumlabelfont}[1]{#1}
\newlength{\thelabelsep}
\setlist{labelsep=\thelabelsep}
\setlist[enumerate,1]{font=\enumlabelfont,label=(\enumlabelformat*),leftmargin=2.5em}
\setlist[itemize]{leftmargin=2.5em,label=$-$}

\newcounter{inlineenum}
\renewcommand{\theinlineenum}{\enumlabelformat{inlineenum}}

\let\epsilon\varepsilon
\let\phi\varphi

\makeatletter
\let\save@mathaccent\mathaccent
\newcommand*\if@single[3]{%
  \setbox0\hbox{${\mathaccent"0362{#1}}^H$}%
  \setbox2\hbox{${\mathaccent"0362{\kern0pt#1}}^H$}%
  \ifdim\ht0=\ht2 #3\else #2\fi
  }
\newcommand*\rel@kern[1]{\kern#1\dimexpr\macc@kerna}
\newcommand*\widebar[1]{\@ifnextchar^{{\wide@bar{#1}{0}}}{\wide@bar{#1}{1}}}
\newcommand*\wide@bar[2]{\if@single{#1}{\wide@bar@{#1}{#2}{1}}{\wide@bar@{#1}{#2}{2}}}
\newcommand*\wide@bar@[3]{%
  \begingroup
  \def\mathaccent##1##2{%
    \let\mathaccent\save@mathaccent
    \if#32 \let\macc@nucleus\first@char \fi
    \setbox\z@\hbox{$\macc@style{\macc@nucleus}_{}$}%
    \setbox\tw@\hbox{$\macc@style{\macc@nucleus}{}_{}$}%
    \dimen@\wd\tw@
    \advance\dimen@-\wd\z@
    \divide\dimen@ 3
    \@tempdima\wd\tw@
    \advance\@tempdima-\scriptspace
    \divide\@tempdima 10
    \advance\dimen@-\@tempdima
    \ifdim\dimen@>\z@ \dimen@0pt\fi
    \rel@kern{0.6}\kern-\dimen@
    \if#31
      \overline{\rel@kern{-0.6}\kern\dimen@\macc@nucleus\rel@kern{0.4}\kern\dimen@}%
      \advance\dimen@0.4\dimexpr\macc@kerna
      \let\final@kern#2%
      \ifdim\dimen@<\z@ \let\final@kern1\fi
      \if\final@kern1 \kern-\dimen@\fi
    \else
      \overline{\rel@kern{-0.6}\kern\dimen@#1}%
    \fi
  }%
  \macc@depth\@ne
  \let\math@bgroup\@empty \let\math@egroup\macc@set@skewchar
  \mathsurround\z@ \frozen@everymath{\mathgroup\macc@group\relax}%
  \macc@set@skewchar\relax
  \let\mathaccentV\macc@nested@a
  \if#31
    \macc@nested@a\relax111{#1}%
  \else
    \def\gobble@till@marker##1\endmarker{}%
    \futurelet\first@char\gobble@till@marker#1\endmarker
    \ifcat\noexpand\first@char A\else
      \def\first@char{}%
    \fi
    \macc@nested@a\relax111{\first@char}%
  \fi
  \endgroup
}
\makeatother

\newcommand{\pd}{\partial}

\title{Examples of cosmological spacetimes without \linebreak CMC Cauchy surfaces}

\author{Eric Ling\footnote{Copenhagen Centre for Geometry and Topology (GeoTop), Department of Mathematical Sciences, University of Copenhagen, DK-2100 Copenhagen, Denmark, el@math.ku.dk}\\Argam Ohanyan\footnote{Department of Mathematics, University of Vienna, Oskar-Morgenstern-Platz 1, 1090 Wien, Austria, \newline argam.ohanyan@univie.ac.at}
}
\begin{document}

\date{\today}


\maketitle

\begin{center} \textit{dedicated to the late Robert Bartnik} \end{center}

\begin{abstract}
CMC (constant mean curvature) Cauchy surfaces play an important role in mathematical relativity as finding solutions to the vacuum Einstein constraint equations is made much simpler by assuming CMC initial data. However,  in \cite{bartnik1988remarks} Bartnik constructed a cosmological spacetime without a CMC  Cauchy surface whose spatial topology is the connected sum of two three-dimensional tori. Similarly, in \cite{chrusciel2005initial}, Chru{\'s}ciel, Isenberg, and Pollack constructed a vacuum cosmological spacetime without CMC Cauchy surfaces whose spatial topology is also the connected sum of two tori. In this article, we enlarge the known number of spatial topologies for cosmological spacetimes without CMC Cauchy surfaces by generalizing Bartnik's construction. Specifically, we show that there are cosmological spacetimes without CMC Cauchy surfaces whose spatial topologies are the connected sum of any compact Euclidean or hyperbolic three-manifold with any another compact Euclidean or hyperbolic three-manifold. Analogous examples in higher spacetime dimensions are also possible. We work with the Tolman-Bondi class of metrics and prove gluing results for variable marginal conditions, which allows for smooth gluing of Schwarzschild to FLRW models.
\vspace{1em}

\noindent
\emph{Keywords:} Cosmological spacetimes, CMC Cauchy surfaces, Tolman-Bondi metrics
\medskip

\noindent
\emph{MSC2020:} 83C20, 53B30, 53C50

\end{abstract}
\newpage
\tableofcontents
\newpage

\section{Introduction}\label{sec:intro}


In this section, we present the sources of interest for ``no-CMC" cosmological spacetimes, i.e.,\ cosmological spacetimes without any CMC Cauchy surfaces. The flow of the presentation follows \cite{galloway2019existence}, where we refer to for many more details, especially regarding historical developments. Let us begin by recalling the cosmological version of the Hawking--Penrose singularity theorem.\footnote{The general form of the Hawking--Penrose theorem makes significantly fewer assumptions on the causality of the spacetime, see e.g.,\ \cite[Thm.\ 12.47]{BeemEhrlich}.}

\begin{Theorem}[Cosmological Hawking--Penrose singularity theorem]
Let $(M,g)$ be a globally hyperbolic spacetime, with compact Cauchy surfaces, satisfying the strong energy condition. (Such spacetimes are sometimes called \emph{cosmological}.) If $(M,g)$ satisfies the generic condition, i.e.,\ along every inextendible causal geodesic $\gamma$ there exists a parameter $t$ such that the tidal force operator 
\begin{align*}
    R: T_{\gamma(t)}M \to T_{\gamma(t)}M, \quad v \mapsto R(v,\gamma'(t))\gamma'(t)
\end{align*}
is not identically zero, then there are incomplete causal geodesics in $M$.
\end{Theorem}

Observe that the assumption of genericity cannot be dropped: Indeed, take $(S,h)$ to be any compact Riemannian manifold with $\Ric \geq 0$, then $(\R \times S, -dt^2 + h)$ satisfies all of the assumptions except for genericity, and is causally geodesically complete. In 1988, Bartnik \cite{bartnik1988remarks} conjectured that this theorem is rigid in the genericity assumption, i.e.,\ that Lorentzian products are the only nongeneric counterexamples.
More precisely, the conjecture states the following:

\begin{Conjecture}[Bartnik's splitting conjecture]
\label{Conjecture: Bartniksplitting}
Let $(M,g)$ be a globally hyperbolic spacetime with compact Cauchy surfaces satisfying the strong energy condition. If $(M,g)$ is timelike geodesically complete, then $(M,g)$ splits isometrically as a product $(\R \times S, -dt^2 + h)$, where $(S,h)$ is a compact Riemannian manifold.
\end{Conjecture}

While this conjecture has been proven over the years under various additional assumptions (we refer to \cite{galloway2019existence} for a detailed discussion of these developments), the version as stated remains open to this day and is one of the most significant open problems in mathematical General Relativity.

In connection with Conjecture \ref{Conjecture: Bartniksplitting}, Bartnik proved the following:

\begin{Theorem}\cite{bartnik1988remarks} Let $(M,g)$ be a globally hyperbolic spacetime, with compact Cauchy surfaces, satisfying the strong energy condition. If $(M,g)$ is timelike geodesically complete, then $(M,g)$ splits isometrically as a product $(\R \times S,-dt^2 + h)$ if and only if there exists a constant mean curvature (CMC) Cauchy surface in $M$.
\end{Theorem}

In \cite{dilts2017spacetimes}, Dilts and Holst review the issue of the existence of CMC Cauchy surfaces in globally hyperbolic spacetimes with compact Cauchy surfaces and raise the question: when do spacetimes have CMC Cauchy surfaces? Motivated by this question, in \cite{galloway2018existence},  Galloway and the first named author proved the existence of a CMC Cauchy surface under the assumptions of compact Cauchy surfaces, future timelike geodesic completeness, and nonpositive timelike sectional curvature. Clearly the latter assumption implies the strong energy condition. Therefore they proposed the following conjecture:

\begin{Conjecture}\cite{galloway2018existence}
\label{Conjecture: CMCBartnik}
Let $(M,g)$ be a globally hyperbolic, future timelike geodesically complete spacetime with compact Cauchy surfaces satisfying the strong energy condition. Then $M$ contains a CMC Cauchy surface.
\end{Conjecture}

As discussed in \cite{dilts2017spacetimes}, most existence results rely on barrier methods, see \cite{Galloway_Ling_Remarks_CMC} for an application. However, a well-known example of Bartnik \cite{bartnik1988remarks} shows that not all spacetimes with compact Cauchy surfaces satisfying the strong energy condition contain CMC Cauchy surfaces. Specifically, Bartnik constructs a timelike geodesically incomplete spacetime satisfying the remaining assumptions of Conjecture \ref{Conjecture: Bartniksplitting} that does not contain any CMC Cauchy surfaces. The rough procedure is as follows: Consider flat FLRW with $T^3$ spacelike slices, cut out a small ball out of $T^3$ and glue the spatial cylinder of Schwarzschild to it. Then extend Schwarzschild to one half of its maximal extension, and glue a time-inverted copy of the resulting spacetime across one of the event horizons in Schwarzschild. The gluing of Schwarzschild to flat FLRW is done in the framework of Tolman-Bondi metrics, reducing the gluing of spacetimes to the smooth interpolation of functions. Using initial data gluing methods, Chru{\'s}ciel, Isenberg, and Pollack construct a similar (vacuum) example obtained in \cite{chrusciel2005initial}. To our knowledge, these are the only such examples which appear in the literature. The goal of this article is to provide  additional ones, by generalizing Bartnik's spacetime gluing construction.

Our generalization of Bartnik's example is two-fold: First, we extend the gluing of Tolman-Bondi metrics to variable marginal conditions using ODE existence and uniqueness arguments (see Theorem \ref{Proposition: GluingTB}). While certainly interesting, this gluing procedure may result in spacetimes which are not globally hyperbolic due to the smallness of the gluing region. To remedy this, we analyze maximality of the ODE solutions at hand and are able to generalize Bartnik's construction by gluing Schwarzschild to flat and hyperbolic FLRW models with compact spacelike slices in such such a way that the resulting spacetime is globally hyperbolic (see Section \ref{sec: generalizationsofBartnikexample} and Theorem \ref{Theorem: generalizationofBartnikconstruction}). This gives a large class of reasonable spacetimes which do not contain any CMC Cauchy surfaces. Note that if Conjecture \ref{Conjecture: Bartniksplitting} is true, then any such example must be incomplete. While Bartnik's construction as well as our generalizations are manifestly incomplete, that incompleteness is a consequence of how the spacetime was constructed. It would be interesting to find similar constructions where the incompleteness is less obvious and more so a consequence of general principles. Such an example would help to gain insight into Conjecture \ref{Conjecture: Bartniksplitting}. In addition, it would be interesting to determine whether or not the example in \cite{chrusciel2005initial} is timelike geodesically complete.

For convenience, let us state the main results (Theorem \ref{Proposition: GluingTB} and Theorem \ref{Theorem: generalizationofBartnikconstruction}) informally:

\begin{Theorem}
\label{Theorem: Informalgluing}
Let $(M_1 = D_1 \times S^2,g_1)$ and $(M_2 = D_2 \times S^2,g_2)$ be Tolman-Bondi spacetimes satisfying Einstein's equations for dust (with nonnegative energy density), where $g_i$ are of the form
\begin{align}
    g_i = -dt^2 + X_i(t,r)^2 dr^2 + Y_i(t,r)^2 d\Omega^2, \quad 0 < X_i,Y_i \in C^{\infty}(D_i),
\end{align}
$D_i \subset \R^2$ open and connected. If certain monotonicity conditions are satisfied, then these spacetimes may be (smoothly) glued across an $r$-interval (where we write $(t,r) \in D_i$) to a larger Tolman-Bondi spacetime $(M,g)$ which also satisfies Einstein's equations for dust with nonnegative energy density.
\end{Theorem}

\begin{Theorem}
\label{Theorem: informalgeneralizations}
Given compact Riemannian quotients $Q,\Tilde{Q}$ of $\R^3$ or $H^3$, we glue Schwarzschild spacetime to a dust-filled FLRW spacetime with spacelike slices $Q$, then glue the resulting spacetime to a gluing of Schwarzschild to FLRW with spacelike slices $\Tilde{Q}$, but with opposite time orientation, across a Schwarzschild event horizon. The resulting spacetime, $(M,g)$, is globally hyperbolic with has compact Cauchy surfaces of topology $ Q\# \Tilde{Q}$ and satisfies the strong and dominant energy conditions. Moreover, $(M,g)$  does not contain any CMC Cauchy surfaces. Analogous examples can be constructed in higher spacetime dimensions by considering higher-dimensional Schwarzschild and FLRW spacetimes.
\end{Theorem}

The paper is organized as follows: In Section \ref{sec:TBspacetimes}, we discuss important properties of Tolman-Bondi spacetimes, where, for convenience, we give full derivations of results we could only find in older literature. We discuss how to describe Schwarzschild and FLRW spacetimes as Tolman-Bondi spacetimes. The section is rounded off by the general gluing result for Tolman-Bondi spacetimes (Theorem \ref{Proposition: GluingTB}). Section \ref{sec: Bartnikexample} is dedicated to a thorough discussion of Bartnik's construction. We review both proofs of Bartnik as to why the constructed spacetime has no CMC Cauchy surfaces. One is of topological nature, relying on the form of the topology of the spacelike slice of FLRW, while the other is more elementary and Lorentz-geometric, using Hawking's singularity theorem and CMC foliation arguments to achieve a contradiction to the existence of CMCs. We describe in Section \ref{sec: generalizationsofBartnikexample} how to generalize Bartnik's construction by gluing Schwarzschild to flat or hyperbolic FLRW with arbitrary compact Riemannian quotients of $\R^3$ or $H^3$ as spacelike slices. One of our proofs makes use of the positive resolution of the surface subgroup conjecture \cite{kahn2012immersing}. We also describe how to obtain the analogous constructions in higher spacetime dimensions. Finally, we give a conclusion and an outlook on possible relevant future lines of research in Section \ref{sec: conclusionoutlook}.


\subsection{Notation and conventions}

Let us collect here some notation and conventions we will use throughout the paper. $A \subset B$ denotes (not necessarily strict) inclusion. A \textit{spacetime} is a connected, time-oriented Lorentzian manifold whose metric has the signature $(-,+,\dots,+)$. By \textit{Cauchy surface} we will always mean a smooth, spacelike hypersurface that is met uniquely by each future directed $C^1$-causal curve $\gamma:(a,b) \to M$ that is inextendible, i.e.,\ the limits $\lim_{t \to a} \gamma(t)$ and $\lim_{t \to b} \gamma(t)$ do not exist.\footnote{This is the ``correct" notion of smooth, spacelike Cauchy surface, cf.\ \cite[Rem.\ 3.31]{minguzzi2019lorentzian}.} $\Ric$ denotes the Ricci tensor and $R$ denotes the scalar curvature. For functions $f(t,r)$ depending on a time variable $t$ and a radial variable $r$, we will write $\dot{f} = \partial_t f$ and $f' = \partial_r f$. We write $T^n$ for the $n$-dimensional torus and $H^n$ for $n$-dimensional hyperbolic space. Given two smooth manifolds $M_1$ and $M_2$, $M_1 \# M_2$ denotes their connected sum. Given groups $G_1$ and $G_2$, $G_1 * G_2$ denotes their free product. Given any spacelike hypersurface $S$ in a spacetime $M$, $H_S \equiv H(S) = \mathrm{Tr}(\nabla N)$ denotes its (future) mean curvature (where $N$ is its future unit normal). We say a spacelike hypersurface $S$ has constant mean curvature (or is CMC) if $H_S$ is a constant. We say it is maximal if $H_S = 0$. We will write $Pr_i$ for the projection map from a given product space to the $i$-th factor. By a \textit{Riemannian quotient} $Q$ of a Riemannian manifold $N$ we mean a quotient arising from the smooth, free, proper and isometric action of a \textit{discrete} Lie group $\Gamma$ on $N$, i.e.\ $Q = N/\Gamma$, equipped with the unique Riemannian metric such that the quotient map $N \to Q$ is a normal Riemannian covering, in particular $Q$ is locally isometric to $N$ and $\dim N = \dim Q$.

\section{Tolman-Bondi spacetimes}\label{sec:TBspacetimes}


Let us begin with a discussion of Tolman-Bondi metrics, which are a class of metrics in four spacetime dimensions with $S^2$-symmetry and which present the setting for the constructions of cosmological spacetimes without CMC Cauchy surfaces. The advantage in using them for our setting is that they allow for purely analytic gluing constructions. We follow the derivations in \cite{bondi1947spherically} and \cite{eardley1979time}, but write out all of the proofs for convenience.

\medskip

\begin{Definition}[Tolman-Bondi spacetimes]
\label{definition: TBspacetimes}
A \emph{Tolman-Bondi (TB) spacetime} is a four-dimensional Lorentzian manifold $M = D \times S^2$ with metric 
\begin{equation}
\label{eq: TBmetric}
    g \,=\, -dt^2 + X(t,r)^2 dr^2 + Y(t,r)^2 d\Omega^2,
\end{equation}
where $D$ is an open connected subset of $\R^2$, and $X,Y$ are smooth positive functions on $D$. Here (and everywhere else), $d\Omega^2$ is the standard round metric on $S^2$. Moreover, $(M,g)$ is understood to be time-oriented via $\frac{\partial}{\partial t}$.


\end{Definition}

Observe that a Tolman-Bondi spacetime is a warped product of the form $D \times_{Y} S^2$, where $S^2$ is equipped with the usual round metric and $D$ is a two-dimensional spacetime with the Lorentzian metric $-dt^2 + X(t,r)dr^2$.

We will be interested in Tolman-Bondi spacetimes satisfying Einstein's equations for dust:

\begin{Definition}[Dust-filled Tolman-Bondi spacetimes]
We say a Tolman-Bondi spacetime $(M = D \times S^2,g)$ is \textit{dust-filled} if it satisfies the Einstein equations for dust, i.e., there exists $\rho \in C^\infty(D)$ such that
\begin{equation}
\label{eq: EEwithdust}
  G:=  \Ric - \frac{1}{2}R g = 8 \pi \rho \, dt^2.
\end{equation}
\end{Definition}

Recall that a spacetime $(M,g)$ is said to satisfy the \emph{strong energy condition (SEC)} if $\Ric(V,V) \geq0$ for all timelike vector fields $V$ on $M$. It satisfies the \emph{dominant energy condition (DEC)} if $G(V,W) \geq 0 $ for future-directed causal vector fields $V,W$ on $M$.

\begin{Lemma}[Dust-filled TB-spacetimes satisfy SEC and DEC]
If $(M,g)$ is a dust-filled Tolman-Bondi spacetime with $\rho$ nonnegative, then it satisfies the strong and dominant energy conditions.
\begin{proof}
We first show the SEC. Complete $\partial_t$ to an orthonormal frame of $M$ and take the $g$-trace of (\ref{eq: EEwithdust}) with respect to that frame to get
\begin{align*}
    R - 2 R = - 8 \pi \rho,
\end{align*}
thus $R = 8 \pi \rho$. Reinserting this into (\ref{eq: EEwithdust}), we get
\begin{align*}
    \Ric = 8 \pi \rho dt^2 + 4 \pi \rho g.
\end{align*}
Thus, if $V$ is any unit timelike vector field on $M$, the reverse Cauchy-Schwarz inequality together with $\rho \geq 0$ implies
\begin{align*}
    \Ric(V,V) \,&=\, 8 \pi \rho \, g(\nabla t, V)^2 - 4\pi \rho \,\geq\, 8 \pi \rho - 4 \pi \rho \,=\, 4 \pi\rho \geq 0,
\end{align*}
thus establishing the SEC.

Showing the DEC is easier: Let $V$ and $W$ be future-directed causal vectors. Then the component $V^t = dt(V)$ is nonnegative since $V$ is future directed. Likewise $W^t \geq 0$. Therefore $G(V,W) = 8\pi \rho V^t W^t \geq 0$. 
\end{proof}
\end{Lemma}

\medskip

\begin{Proposition}
\label{Proposition: quantitiesindependentoft}
Let $(M,g)$ be a dust-filled Tolman-Bondi spacetime. Then
$\pd_t(\frac{1}{X}Y') = 0$ and $\pd_t(\rho Y' Y^2) = 0$.
\begin{proof}
Let $G = \Ric - \frac{1}{2}R g$ denote the Einstein tensor. Since the Einstein equations hold with dust, we have $G(\pd_t, \pd_r) = 0$. A calculation shows 
\begin{align}
\label{eq: Gdtdr}
G(\pd_t, \pd_r) \,=\, \frac{2}{XY}(\dot{X}Y' - X \dot{Y}').
\end{align}
Therefore $\dot{X}Y' - X\dot{Y}' = 0$ which implies
\[
 X^2\pd_t(Y'/X) \,=\, 0.
\]
Since $X$ is never zero, $\pd_t(Y'/X) = 0$. 

To prove that $\rho Y' Y^2$ is independent of $t$, we investigate $G(\pd_t, \pd_t)$ and $G(\pd_r, \pd_r)$: An explicit calculation yields
\begin{align}
\label{eq: Gdtdt}
G(\pd_t, \pd_t) \,&=\, \frac{1}{X^3 Y^2}\bigg(X^3 \dot{Y}^2 + 2 X^2 \dot{X} Y \dot{Y} + X^3 - 2XY Y'' - X Y^{\prime\, 2} + 2X' YY' \bigg)
\\
\label{eq: Gdrdr}
G(\pd_r, \pd_r) \,&=\, -\frac{1}{Y^2}\bigg(2X^2Y\ddot{Y} + X^2\dot{Y}^2 + X^2 - Y^{\prime\, 2}\bigg).
\end{align}
We introduce the quantity
\begin{align}
\label{eq: defofU}
    S := Y\left(1 + \dot{Y}^2 -  \frac{Y^{\prime\, 2}}{X^2}\right).
\end{align}
It is elementary to check that
\begin{align}
\label{eq: usefulGdtdt}
G(\pd_t, \pd_t)Y' \,&=\, \frac{1}{Y^2}S',
\\
\label{eq: usefulGdrdr}
G(\pd_r, \pd_r) \dot{Y} \,&=\, -\frac{X^2}{Y^2 }\dot{S}.
\end{align}
Now, $\dot{S} = 0$ since $G(\pd_r, \pd_r) = 0$. Therefore $\dot{S}' = 0$, hence
\[
\pd_t \big(Y^2 Y' G(\pd_t, \pd_t)\big) \,=\,0 .
\]
Since $G(\pd_t,\pd_t) = 8 \pi \rho$, the claim follows.
\end{proof}
\end{Proposition}



The next result is a key property of dust-filled Tolman-Bondi spacetimes, as it allows one to view the coefficient functions $X$ and $Y$ as solutions of ODEs.

\begin{Proposition}
\label{Proposition: ODEforY}
    Let $(M,g)$ be a dust-filled Tolman-Bondi spacetime. Then the function $S$ defined in Equation (\ref{eq: defofU}) is independent of $t$ and is an antiderivative of $8\pi \rho Y' Y^2$ (the latter is independent of $t$ due to Proposition \ref{Proposition: quantitiesindependentoft}). Moreover, writing $W:=Y'/X$, it holds that
    \begin{align}
    \label{eq: ODEforY}
        \dot{Y}(t,r)^2 = W(r)^2 - 1 + \frac{1}{Y(t,r)} S(r).
    \end{align}
    \begin{proof}
        Let $S$ be the function introduced in Equation (\ref{eq: defofU}). We have already shown in the proof of Proposition \ref{Proposition: quantitiesindependentoft} that $\dot{S} = 0$ (this is a consequence of Equation (\ref{eq: usefulGdrdr})). By Equation (\ref{eq: usefulGdtdt}), and noting that $G(\partial_t,\partial_t) = 8 \pi \rho $, we get
        \begin{align*}
            S' = 8 \pi \rho Y' Y^2.
        \end{align*}
        Inserting the definition of $S$, we get
        \begin{align*}
            S = Y \left(1+\dot{Y}^2 - \frac{Y'^2}{X^2}\right),
        \end{align*}
        solving for $\dot{Y}^2$ gives the claim.
    \end{proof}
\end{Proposition}

\begin{Definition}[Marginal condition and mass function]
    Let $(M,g)$ be a dust-filled Tolman Bondi spacetime, 
    \begin{align*}
        g = -dt^2 + X^2 dr^2 + Y^2 d\Omega^2.
    \end{align*}
    We call the associated functions $W:=Y' X^{-1}$ and $S:=Y(1+\dot{Y}^2 - Y'^2 X^{-2})$ \textit{marginal condition} and $\textit{mass function}$\footnote{$S/2$ can justifiably be called \textit{mass} of the Tolman-Bondi spacetime (in Schwarzschild: $S \equiv 2m$). We prefer to work with $S$ and have thus chosen the name \textit{mass function} to avoid confusion.}, respectively.
\end{Definition}

Next, we discuss some important examples of Tolman-Bondi spacetimes, namely Schwarzschild and FLRW. These will be the building blocks for constructing cosmological spacetimes without CMC Cauchy surfaces.

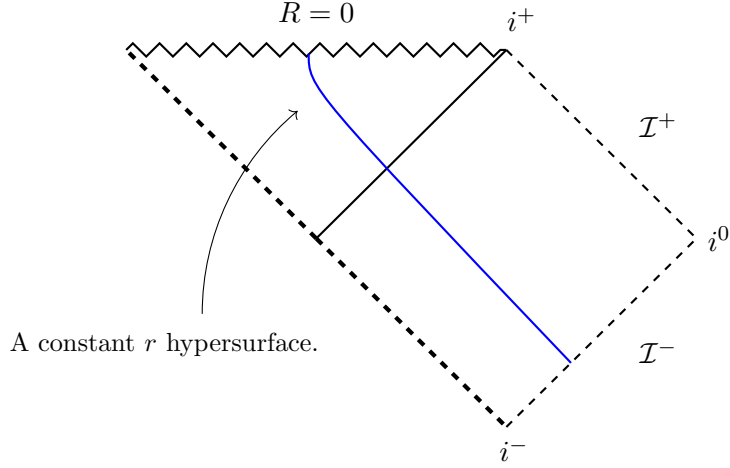
\begin{figure}[h]
\[
\begin{tikzpicture}[scale=1.0]


\draw[snake=zigzag,thick] (-14.5,12) -- (-9.5,12);

\draw[-,dashed, ultra thick] (-12,9.5) -- (-9.5,7);
\draw[-,thick] (-12,9.5) -- (-9.5,12);
\draw[-,dashed, thick] (-9.5,12) -- (-7,9.5);
\draw[-,dashed, thick] (-9.5,7) --(-7,9.5);

\draw[-,dashed, ultra thick] (-12,9.5) -- (-14.5,12);

\node [scale=1.0] at (-7.5,11.00)  {$\mathcal{I}^+$}; 
\node [scale=1.0] at (-7.5,8.00)  {$\mathcal{I}^-$}; 

\node [scale=1.0] at (-12,12.50)
{$R = 0$}; 

\node [scale=1.0] at (-9.3,12.40)
{$i^+$}; 
\node [scale=1.0] at (-9.4,6.70)
{$i^-$}; 

\node [scale=1.0] at (-6.7,9.5)
{$i^0$}; 

\draw [thick, blue] (-8.65,7.85) .. controls (-12.1,11.5).. (-12.1,11.95);

\draw [->] (-13.5,8.5) arc [start angle=180, end angle=130, radius=100pt];
\node at (-14,8.1) [scale = .9] {A constant $r$ hypersurface.};

\end{tikzpicture}
\]
\caption{\small{The Lemaître coordinates $(t,r)$ cover one-half of the maximally extended Schwarzschild spacetime. The hypersurfaces of constant $r$ are timelike and foliate the spacetime; they become null as they approach $\mathcal{I}^-$. As $r \to -\infty$, the hypersurfaces approach the (totally geodesic) null hypersurface represented by the thick dashed lines, i.e., the hypersurface $T = -X$ in Kruskal coordinates. }}
\label{fig: schwarz in Lemaitre coord}
\end{figure}

\begin{Example}[Schwarzschild spacetime]
\label{example: Schwarzschild}
The standard (non-extended) Schwarzschild spacetime of mass $m > 0$ is $(R \neq 0,2m)$
\begin{align*}
    g_S:= - \left(1 - \frac{2m}{R}\right) d\tau^2 + \frac{1}{1-\frac{2m}{R}} dR^2 + R^2 d\Omega^2.
\end{align*}
We introduce Lemaître coordinates $t,r$ via
\begin{align*}
    &dt = d\tau + \sqrt{\frac{2m}{R}} \frac{dR}{1- \frac{2m}{R}},\\
    &dr = d\tau + \sqrt{\frac{R}{2m}} \frac{dR}{1-\frac{2m}{R}}.
\end{align*}
Writing the metric in these coordinates yields
\begin{align*}
    g_S = -dt^2 + \left(\frac{2m}{\frac{3}{2}(r-t)}\right)^{\frac{2}{3}} dr^2 + (2m)^{\frac{2}{3}} \left( \frac{3}{2}(r-t)\right)^{\frac{4}{3}} d\Omega^2.
\end{align*}
This is in Tolman-Bondi form defined on $M:=\{(t,r) \in \R^2 : t < r\} \times S^2$ and describes one half of the maximal analytic extension of the Schwarzschild spacetime. The hypersurfaces $\{r = const\}$ are timelike and, for $r \to -\infty$, converge to the event horizon separating the two halves of the maximal extension (see Figure \ref{fig: schwarz in Lemaitre coord}). For later use, let us note that
\begin{align*}
    &W(r) = 1,\\
    &S(r) = 2m.
\end{align*}
\end{Example}

\begin{Example}[FLRW]
\label{Example: FLRW}
We will now describe the FLRW spacetimes with any of the constant spatial sectional curvatures $k=+1,0,-1$ as Tolman-Bondi metrics. We will refer to these as \textit{spherical, flat and hyperbolic} FLRW, respectively. We refer to \cite[Sec.\ 12.5]{o1983semi} for more details than what is presented below. Note that we work with negative time intervals for technical reasons, as this will later allow us to glue FLRW spacetimes to Schwarzschild using Theorem \ref{Proposition: GluingTB} (resp.\ Corollary \ref{Corollary: GluingTBtoSchwarzschild}) in conjunction with Propositions \ref{Proposition: ExplicitTB} and \ref{Proposition: W > 1 full description}.
\begin{enumerate}
    \item Flat FLRW: For any parameter $\mathcal{M} > 0$, the spacetime $(-\infty,0) \times_{a_f} \R^3$, with
    \begin{align*}
        a_f(t):=(6 \pi \mathcal{M})^{\frac{1}{3}} t^{\frac{2}{3}},
    \end{align*}
    satisfies Einstein's equations for dust, with energy density $\rho(t) = \mathcal{M}a_f(t)^{-3}$, where $\mathcal{M}$ is the Friedmann mass parameter. Writing the $3$-dimensional Euclidean metric in spherical coordinates as $dr^2 + r^2 d\Omega^2$, the metric for flat FLRW can be written in Tolman-Bondi form
    \begin{align*}
        g_f = -dt^2 + a_f(t)^2 dr^2 + a_f(t)^2 r^2 d\Omega^2
    \end{align*}
    on the manifold $(-\infty,0) \times (0,\infty) \times S^2$.
    We calculate $W$ and $S$ for this metric:
    \begin{align*}
        &W(r) = 1,\\
        &S(r) = \frac{8\pi \mathcal{M}}{3} r^3.
    \end{align*}
    \item Hyperbolic FLRW: Here, the spacetime is $(-\infty,0) \times_{a_h} H^3$. The function $a_h$ cannot be written explicitly, but it satisfies Friedmann's equation
    \begin{align*}
        \dot{a}_h^2 - 1 = \frac{8 \pi \mathcal{M}}{3 a_h}
    \end{align*}
    and can be written in dependence on a parameter $\eta < 0$ (together with time) as
    \begin{align*}
        t(\eta) = \frac{4 \pi \mathcal{M}}{3} (\sinh(\eta) - \eta),\\ a_h(\eta) = \frac{4\pi \mathcal{M}}{3}(\cosh(\eta) - 1).
    \end{align*}
    Writing the hyperbolic metric on $H^3$ as $dr^2 + \sinh(r)^2 d\Omega^2$ (on $(0,\infty) \times S^2$), the hyperbolic FLRW metric can be expressed on the manifold $(-\infty,0) \times (0,\infty) \times S^2$ in Tolman-Bondi form as
    \begin{align*}
        g_h = -dt^2 + a_h(t)^2 dr^2 + a_h(t)^2 \sinh(r)^2 d\Omega^2.
    \end{align*}
    The energy density is again $\rho(t) = \mathcal{M} a_h(t)^{-3}$ and moreover
    \begin{align*}
        &W(r) = \cosh(r),\\
        &S(r) =  \frac{8 \pi \mathcal{M}}{3} \sinh(r)^3.
    \end{align*}
    \item Spherical FLRW: Similarly to before, consider $\left(-\frac{8 \pi^2 \mathcal{M}}{3},0\right) \times_{a_s} S^3$, where $a_s$ satisfies
    \begin{align*}
        \dot{a_s}^2 + 1 = \frac{8 \pi \mathcal{M}}{3 a_s}. 
    \end{align*}
    As in the hyperbolic case, $a_s$ cannot be written explicitly, but can be described in dependence on a parameter $\theta \in (-2\pi, 0)$ (together with time) via
    \begin{align*}
        &t(\theta) = \frac{4\pi \mathcal{M}}{3} ( \theta - \sin(\theta)),\\
        &a_s(\theta) = \frac{4 \pi \mathcal{M}}{3} (1 - \cos(\theta)).
    \end{align*}
    Writing the metric on $S^3$ as $dr^2 + \sin(r)^2 d\Omega^2$ (on $(0,\pi) \times S^2$), the spherical FLRW metric can be written on the manifold $\left(-\frac{8 \pi^2 \mathcal{M}}{3},0\right) \times (0,\pi) \times S^2$ as
    \begin{align*}
        g_s = -dt^2 + a_s(t)^2 dr^2 + a_s(t)^2 \sin(r)^2 d\Omega^2.
    \end{align*}
    Here, $\rho(t) = \mathcal{M} a_s(t)^{-3}$ and
    \begin{align*}
        &W(r) = \cos(r),\\
        &S(r) = \frac{8 \pi \mathcal{M}}{3} \sin(r)^3.
    \end{align*}
\end{enumerate}

\end{Example}

The next result shows that, under certain monotonicity conditions, we can always glue dust-filled Tolman-Bondi spacetimes across intervals of $r$ by solving the ODE for $Y$ given in Proposition \ref{Proposition: ODEforY}.

\begin{Theorem}[Gluing Tolman-Bondi spacetimes]
\label{Proposition: GluingTB}
Let $(M_1 = D_1 \times S^2, g_1)$ and $(M_2 = D_2 \times S^2, g_2)$ be dust-filled Tolman-Bondi spacetimes with nonnegative energy densities $\rho_1,\rho_2$. We write $Pr_2(D_i) = (r_i^-,r_i^+)$, $i=1,2$. Suppose $r_1 \in (r_1^-,r_1^+)$ and $r_2 \in (r_2^-,r_2^+)$, $r_1 < r_2$. Moreover, suppose there exists $t_0$ such that $(t_0,r_i) \in D_i$, such that either both $\dot{Y}_1(t_0,r_i) > 0$ and $\dot{Y}_2(t_0,r_i) > 0$ or both $\dot{Y}(t_0,r_1) < 0$ and $\dot{Y}(t_0,r_2) < 0$, and suppose that the following technical assumption is satisfied:
There exist smooth functions $W,S,r \mapsto Y_r(t_0) \in C^{\infty}((r_1^-,r_2^+))$ agreeing with $W_i,S_i,r \mapsto Y_i(t_0,r)$ on their corresponding domains with $S$ monotonically increasing, $r \mapsto Y_r(t_0)$ strictly monotonically increasing, $W$ positive, such that on $[r_1,r_2]$,
    \begin{align*}
        W(r)^2 - 1 + \frac{S(r)}{Y_r(t_0)} > 0.
    \end{align*}
Then there exists a dust-filled Tolman-Bondi spacetime $(M=D \times S^2,g)$ with a nonnegative energy density $\rho$ such that $Pr_2(D) = (r_1^-,r_2^+)$, $g = g_1$ for $r \leq r_1$ and $g = g_2$ for $r \geq r_2$. 
\begin{proof}
    Let us assume for definiteness that $\dot{Y}_1(t_0,r_1) < 0$ and $\dot{Y}_2(t_0,r_2) < 0$. We are interested in solving the following ODE initial value problem for fixed $r \in [r_1,r_2]$:
    \begin{align*}
        \begin{cases}\dot{Y}(t,r) = -\sqrt{W(r)^2 - 1 + \frac{S(r)}{Y(t,r)}},\\
        Y(t_0,r) = Y_r(t_0).\end{cases}
    \end{align*}
    First, the fact that $\dot{Y}_i(\cdot,r_i) < 0$ implies that for all $t$ close to $t_0$,
    \begin{align*}
        \dot{Y}_i(t,r_i) = -\sqrt{W_i(r_i)^2 - 1 + \frac{S_i(r_i)}{Y_i(t,r_i)}}.
    \end{align*}
    
    Let $S,W,Y_r(t_0)$ be as in the assumption. Due to positivity, the right-hand side of the initial value problem in question depends smoothly on $Y(\cdot,r)$, for every fixed $r \in [r_1,r_2]$, hence admits a unique solution $Y(t,r)$ on a maximal domain of definition $(t_r^-,t_r^+)$. We may extract a common domain of definition $(t^-,t^+)$ such that all $Y(t,r)$, $r \in [r_1,r_2]$, are defined for $t \in (t^-,t^+)$. Moreover, since the initial data $r \mapsto Y_r(t_0)$ depends smoothly on $r$, so does the solution. We may assume that $(t^-,t^+)$ is small enough so that $Y, Y' > 0, \dot{Y}  < 0$ for all $(t,r) \in (t^-,t^+) \times [r_1, r_2]$.
    
    Let us now construct the spacetime: We set $D:=\tilde{D}_1 \cup ((t^-,t^+) \times [r_1,r_2]) \cup \tilde{D}_2$, where
    \begin{align*}
        &\Tilde{D}_1:=\{(t,r) \in D_1 : r < r_1\},\\
        &\Tilde{D}_2:=\{(t,r) \in D_2 : r > r_2\}.
    \end{align*}
    
    For $(t,r) \in (t^-,t^+) \times [r_1,r_2]$, we let $Y$ be the solution obtained above and $X:=Y'/W$. Then $(t,r) \in D \mapsto X(t,r),Y(t,r)$ are positive, smooth functions on $D$. Moreover, by construction, the corresponding Tolman-Bondi metric is dust-filled with energy density given for $(t,r) \in (t^-,t^+) \times [r_1,r_2]$ by
    \begin{align*}
        \rho:=\frac{S'(r)}{8 \pi Y^2 Y'} \geq 0.  
    \end{align*}
    To elucidate, equations \eqref{eq: Gdtdr} - \eqref{eq: usefulGdrdr} in the proof of Proposition \ref{Proposition: quantitiesindependentoft} are true for any Tolman-Bondi metric. Therefore $G(\pd_t, \pd_r) = 0$ since $X = Y'/W$ and $G(\pd_r, \pd_r) = 0$ since $\dot{S} = 0$. Thus the Tolman-Bondi metric is dust-filled.
\end{proof}
\end{Theorem}

\begin{Remark}
\label{Remark: gluingeasyWcase}
    In many cases, it is elementary to satisfy the technical assumption of Theorem \ref{Proposition: GluingTB}: Indeed, if
    \begin{enumerate}
        \item either $\dot{Y}_1(t_0,r_1) > 0$ and $\dot{Y}_2(t_0,r_2) > 0$ or $\dot{Y}_1(t_0,r_1) < 0$ and $\dot{Y}_2(t_0,r_2) < 0$,
        \item $0 < S_1(r_1) \leq S_2(r_2)$,
        \item $Y_1(t_0,r_1) < Y_2(t_0,r_2), \quad Y_1'(t_0,r_1) > 0, \quad  Y_2'(t_0,r_2) > 0,$
        \item $W_1(r_1)^2 \geq 1$ and $W_2(r_2)^2 \geq 1$, $W_1 > 0$ and $W_2 > 0$,
    \end{enumerate}
    then one can simply choose monotonically increasing (resp.\ strictly monotonically increasing) connecting functions $S,Y_r(t_0)$ as well as any interpolating function $W$ which increases resp.\ decreases from $W_1(r_1)$ to $W_2(r_2)$ (if $W_1(r_1) \leq W_2(r_2)$ resp.\ $W_1(r_1) \geq W_2(r_2)$). 
\end{Remark}

\begin{Remark}[Gluing vacuum TB-spacetimes]
Note that if $(M_1,g_1)$ and $(M_2,g_2)$ are vacuum Tolman-Bondi spacetimes, i.e.,\ $\rho_1 = 0$ and $\rho_2 = 0$, then $S_1,S_2$ are constants. If $S_1 = S_2$ and the remaining assumptions of Theorem \ref{Proposition: GluingTB} are satisfied, the proof shows that, by choosing the connecting smooth function $S$ to be that same constant, the glued spacetime can be constructed to be vacuum. 
\end{Remark}

Let us note the following useful consequence of Theorem \ref{Proposition: GluingTB} regarding the gluing of Schwarzschild spacetime to (fairly) arbitrary Tolman-Bondi spacetimes.

\begin{Corollary}[Gluing TB-spacetimes to Schwarzschild]
\label{Corollary: GluingTBtoSchwarzschild}
Let $(M_1,g_1)$ be Schwarzschild spacetime of mass $m$ in Lemaître coordinates as described in Example \ref{example: Schwarzschild}. Let $(M_2 = D_2 \times S^2,g_2)$ be any dust-filled Tolman-Bondi spacetime with nonnegative energy density $\rho_2$. Let $r_1 \in \R$. Suppose there is an $r_2 > r_1$ and $t_0 < r_1$ such that $(t_0, r_2) \in D_2$. Moreover, suppose the following technical assumptions are satisfied:
\begin{enumerate}
    \item $\dot{Y}_2(t_0,r_2) < 0$.
    \item $2m \leq S_2(r_2)$.
    \item $(2m)^{\frac{1}{3}} \left(\frac{3}{2}(r_1 - t_0)\right)^{\frac{2}{3}} < Y_2(t_0,r_2)$, $Y_2'(t_0,r_2) > 0$.
\end{enumerate}
Then there exists a  dust-filled Tolman-Bondi spacetime $(M=D \times S^2,g)$ with a nonnegative energy density $\rho$ such that $Pr_2(D) = (-\infty, \sup Pr_2(D_2))$, $g|_{(-\infty,r_1]} = g_1$, and $g|_{[r_2,\sup Pr_2(D_2))} = g_2$. 
\begin{proof}
    We need to verify the technical condition in Theorem \ref{Proposition: GluingTB}. Indeed, pick any smooth, monotonically increasing connector $S$ for $S_1,S_2$ and strictly monotonically increasing connector $Y_r(t_0)$ for $Y_1(t_0,\cdot)$ and $Y_2(t_0,\cdot)$ on the interval $[r_1,r_2]$. Note that $W_1 \equiv 1$, and $W_2(r_2) = \frac{Y_2'(t_0,r_2)}{X_2(t_0,r_2)} > 0$ by assumption. If $W_2(r_2) \geq 1$, we are done by Remark \ref{Remark: gluingeasyWcase}, so suppose $0 < W_2(r_2) < 1$. Moreover,
    \begin{align*}
        W_2(r_2)^2 - 1 + \frac{S_2(r_2)}{Y_2(t_0,r_2)} > 0
    \end{align*}
    since $\dot{Y}_2(t_0,r_2) < 0$. Let $\delta > 0$ be such that
    \begin{align*}
        \frac{S_2(r_2)}{Y_2(t_0,r_2)} > 1 - W_2(r_2)^2 + \delta.
    \end{align*}
    Let $\varepsilon > 0$ be such that
    \begin{align*}
        \frac{S(r)}{Y_r(t_0)} > 1 - W_2(r_2)^2 + \frac{\delta}{2}
    \end{align*}
    on $[r_2-\varepsilon,r_2]$. Now let $W$ be a smooth function connecting $W_1$ to $W_2$ as follows: $W \equiv 1$ on $[r_1,r_2-\varepsilon]$ and smoothly and monotonically decreases on $[r_2-\varepsilon,r_2]$ to $W_2(r_2)$. Then on $[r_2 -\varepsilon,r_2]$,
    \begin{align*}
        \frac{S(r)}{Y_r(t_0)} > 1 - W_2(r_2)^2 + \frac{\delta}{2} \geq 1 - W(r)^2 + \frac{\delta}{2}.
    \end{align*}
    Since $W = 1$ on $[r_1,r_2 - \varepsilon]$, we have thus shown that the technical condition in Theorem \ref{Proposition: GluingTB} is satisfied. This concludes the proof.
\end{proof}
\end{Corollary}

\begin{Remark}
While the preceding results show that the gluing of Tolman-Bondi spacetimes can in principle be done under fairly general circumstances, the resulting construction will in general not be globally hyperbolic since the glued region may be very small. This can be seen in examples where a Tolman-Bondi spacetime is glued to Schwarzschild (cf.\ Corollary \ref{Corollary: GluingTBtoSchwarzschild}): If the glued region does not extend to past timelike infinity $i^-$ (i.e.\ if $t_- \neq - \infty$ in the proof of Theorem \ref{Proposition: GluingTB}), then one may find timelike curves emanating from $i^-$ which are unable to cross into the glued region. Thus the glued spacetime does not contain any Cauchy hypersurfaces.

To ensure global hyperbolicity, an investigation into the maximal interval of existence of the ODE in Proposition \ref{Proposition: ODEforY} is necessary.
\end{Remark}

\section{Bartnik's ``no-CMC" example}\label{sec: Bartnikexample}


In this Section, we present the cosmological spacetime without CMC Cauchy surfaces constructed by Bartnik \cite{bartnik1988remarks}. An essential ingredient of the construction is the gluing Schwarzschild spacetime to flat FLRW. In this case, the ODE given in Proposition \ref{eq: ODEforY} may be solved explicitly, allowing for a simple gluing argument in terms of functions.

\begin{Proposition}[Explicit TB-metrics]
\label{Proposition: ExplicitTB}
Let $( D \times S^2,g)$ be a dust-filled Tolman-Bondi spacetime with marginal condition $W(r) \equiv 1$ and suppose the mass function is always positive, $S > 0$. 
Then, writing $M(r)=S(r)/2$, there exists a smooth function $t_0(r)$ such that\footnote{We use $M$ instead of $S$ here to be consistent with the sources \cite{bondi1947spherically}, \cite{eardley1979time} and \cite{bartnik1988remarks}. The function $t_0(r)$ is an integration constant depending on $r$ which arises from solving the ODE for $Y$.}
\begin{align}
\label{eq: W=1explicitX}
&X(t,r) = \frac{M'(r) (t_0(r) - t) + 2M(r)t_0'(r)}{[6 M(r)^2 (t_0(r) - 
t)]^{\frac{1}{3}}},\\
\label{eq: W=1explicitY}
&Y(t,r) = \left( \frac{9}{2} M(r) (t_0(r) - t)^2\right)^{\frac{1}{3}}.
\end{align}
Moreover, the metric can be extended smoothly to $\tilde{D} \times S^2$, where $\tilde{D} = \{(t,r) \in \R^2 : r \in \{M > 0\}, t \in (-\infty, t_0(r))\}$, provided that
\begin{align}
\label{eq: monotonicityMt_0}
    \frac{d}{dr} \left( M(r) (t_0(r) - t)^2 \right) > 0.
\end{align}
Conversely, given smooth functions $M(r)$ and $t_0(r)$ defined on $I \subset \R$ with $M > 0$ satisfying $(\ref{eq: monotonicityMt_0})$ (this is in particular the case if $t_0$ is strictly monotonically increasing and $M$ is monotonically increasing, or vice versa), Equations (\ref{eq: W=1explicitX}) and (\ref{eq: W=1explicitY}) define a Tolman-Bondi spacetime on $\tilde{D} \times S^2$, where $\tilde{D}$ is given as before ($I = \{M > 0\}$).
\begin{proof}
    Since $W = 1$ and $S > 0$, by Equation (\ref{eq: ODEforY}) $\dot{Y}$ is either always positive or always negative, w.l.o.g.\ let us assume that $\dot{Y} < 0$\footnote{This is the case for Schwarzschild and flat FLRW, cf.\ Example \ref{example: Schwarzschild} and Example \ref{Example: FLRW}.}. Then, by the same reference,
    \begin{align*}
        \dot{Y}(t,r) = - \sqrt{\frac{S(r)}{Y(t,r)}}.
    \end{align*}
    By assumption $S(r) > 0$, so fixing some $s_0$ such that $(s_0,r) \in D$, $Y(t,r)$ has to agree with the unique solution $\tilde{Y}(t,r)$ of the following $r$-parameter family of ODEs:
    \begin{align*}
        \begin{cases}\dot{\tilde{Y}}(t,r) = -\sqrt{\frac{S(r)}{\tilde{Y}(t,r)}},\\
        \tilde{Y}(s_0,r) = Y(s_0,r).\end{cases}
    \end{align*}
    This ODE can be integrated explicitly and yields Equation (\ref{eq: W=1explicitY}) with $M:=S/2$ and
    \begin{align*}
        t_0(r) = s_0 + \frac{\sqrt{2}}{3\sqrt{M(r)}} Y(s_0,r)^{\frac{3}{2}}.
    \end{align*}
    Evidently, the maximal existence interval for $Y(t,r)$ is $t \in (-\infty,t_0(r))$. The condition (\ref{eq: monotonicityMt_0}) guarantees that $X = Y' > 0$ on $\tilde{D}$. 
\end{proof}
\end{Proposition}

Let us now describe the construction: Let $(M_1,g_1)$ be the following portion of Schwarzschild spacetime in Lemaître coordinates with mass $m=1$, as described in Example \ref{example: Schwarzschild}: $M_1 = \{(t,r) \in \R^2: t < r\leq 1- \varepsilon\} \times S^2$ (for some small $\varepsilon > 0$). Since Schwarzschild has marginal condition $W = 1$, the functions $X,Y$ are given in terms of functions $M(r)$ and $t_0(r)$ as described in Proposition \ref{Proposition: ExplicitTB}. It is easily checked that
\begin{align}
\label{eq: Mt0forSchwarzschild}
M(r) = 1, \quad t_0(r)  = r.
\end{align}
Next, consider $T^3$ as a Riemannian quotient of $\R^3$ with the group action induced by $3\mathbb{Z}^3$, hence we can consider $(-\infty,1) \times_{a_f} T^3$ as a Lorentzian quotient of flat FLRW $(-\infty, 1) \times_{a_f} \R^3$ with Friedmann mass parameter $\mathcal{M} = \frac{3}{4\pi}$. Note that we shift the upper limit of the time parameter by $1$ for monotonicity reasons which will become clear in a moment. Let $x_0 \in T^3$ be arbitrary and $0 < 2\varepsilon < 1$. Then $B^{T^3}_{1+2\varepsilon}(x_0)$ is isometric to $B^{\R^3}_{1+2\varepsilon}(y_0)\subset \R^3$ (for any $y_0$ in the preimage of $x_0$ under the quotient map), hence the spacetime portion $(-\infty,1) \times_{a_f} (B^{T^3}_{1+2\varepsilon}(x_0) \setminus B^{T^3}_{1+\varepsilon}(x_0))$ can be written in Tolman-Bondi form, identical to flat FLRW. Let $(M_2,g_2)$ be the spacetime $(-\infty,1) \times_{a_f} (T^3 \setminus B^{T^3}_{1+\varepsilon}(x_0))$. Since flat FLRW also has marginal condition $W = 1$, it is also given explicitly according to Proposition \ref{Proposition: ExplicitTB} with functions 
\begin{align}
    M(r) = r^3, \quad t_0(r) = 1.
\end{align}
We now glue $(M_1,g_1)$ to $(M_2,g_2)$ across the $r$-interval $[1-\varepsilon,1+\varepsilon]$: Indeed, due to the explicit form of the functions $M,t_0$ in the two spacetimes, we may connect them smoothly across $[1-\varepsilon,1+\varepsilon]$ while satisfying (on $(1-\varepsilon,1+\varepsilon)$)\footnote{Bartnik \cite{bartnik1988remarks} demands in addition that $t_0'' \leq 0$, $M'' \geq 0$ and $t_0^2 + M^2 > 0$. The latter is automatic, and the convexity/concavity conditions (while certainly achievable) do not appear to be necessary.}
\begin{align}
    t_0' > 0 \text{ and } M' > 0.
\end{align}
(This is the reason for shifting the time parameter of FLRW by $1$, so that $t_0' > 0$ can be achieved). As we argued in the proof of Proposition \ref{Proposition: ExplicitTB}, $t_0' > 0$ and $M' > 0$ imply $Y,Y' = X > 0$, as well as $\rho = M'/4\pi Y^2 Y' \geq 0$. We denote the resulting spacetime by $(M^+,g^+)$.

\begin{figure}[h]
\[
\begin{tikzpicture}[scale=1.0]


\draw[->, thick] (-3,-3) --(-0.5,-3);
\draw[->, thick] (-0.5, -3) -- (0.5, -3);
\draw[thick] (0.5, -3) -- (3,-3);

\draw[->, thick] (-3,3) --(-0.5,3);
\draw[->, thick] (-0.5, 3) -- (0.5, 3);
\draw[-, thick] (0.5, 3) -- (3,3);

\draw[->, thick] (-3,-3) -- (-3,0);
\draw[-, thick] (-3,0) -- (-3,3);

\draw[->, thick] (3,-3) -- (3,0);
\draw[-, thick] (3,0) -- (3,3);


\draw[ultra thick] (0,0) circle [radius=2];
\draw[ultra thick] (0,0) circle [radius=1.5];
\draw[thick] (0,0) circle [radius =.05];

\fill[gray!40,even odd rule] (0,0) circle (2) (0,0) circle (1.5);

\end{tikzpicture}
\]
\caption{\small{A two-dimensional spatial cross section of $(M^+, g^+)$. The inner and outer circles are the spheres $r = 1 - \varepsilon$ and $r = 1 + \varepsilon$, respectively. For $r < 1-\varepsilon$, the spacetime is exactly given by the Schwarzschild spacetime. For $r > 1 +\varepsilon$, the spacetime is exactly given by a $k = 0$ FLRW dust spacetime; at large values $r$ ceases to be a coordinate since we identify the sides. The shaded region between the spheres represents the gluing region. The point at the center represents $r = -\infty$. }}
\label{fig: schw to flrw glue}
\end{figure}
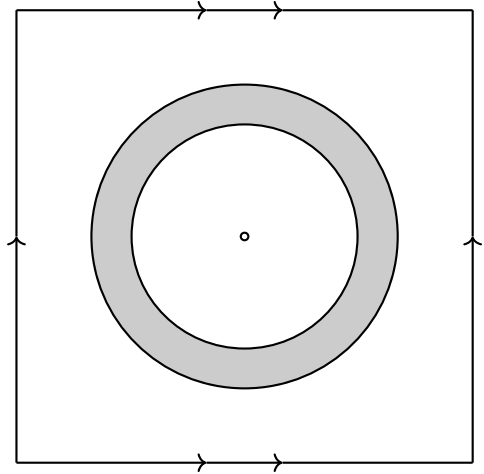

\begin{Lemma}
    $(M^+,g^+)$ is globally hyperbolic.
\end{Lemma}

We omit the proof of this Lemma, since it is very similar to the one we give in the generalization to the hyperbolic case in Lemma \ref{Lemma: M+g+ glob hyp in hyp case}.

Now let $(M^-,g^-)$ be a time-reversed copy of $(M^+,g^+)$. We extend them both to $r = -\infty$ (which corresponds to extending Schwarzschild to its event horizon) and glue them there\footnote{One could have also started out with two sides of the Schwarzschild horizon in the Kruskal extension and then glued FLRW to both of them in Lemaître and time-reversed Lemaître coordinates, it would have amounted to the same construction.}. The resulting Lorentzian manifold $(M,g)$ is smooth and time-orientable due to the properties of the maximal extension of Schwarzschild. Observe that there exists a time-inverting isometry $\phi:(M,g) \to (M,g)$ mapping $(M^+,g^+)$ to $(M^-,g^-)$ and vice versa. Now, since both $(M^+,g^+)$ and $(M^-,g^-)$ are globally hyperbolic, so is $(M,g)$ (this can be seen by looking at the corresponding Penrose diagram, see \cite[Fig.\ 2]{bartnik1988remarks}). A Cauchy surface can be constructed by connecting the bifurcate horizon smoothly to one of the spatial slices in the each of the FLRW regions. \cite[Cor.\ 14.54]{o1983semi} can be used to show that this is in fact a Cauchy surface.

We will give the two proofs of Bartnik as to why $(M,g)$ has no CMC Cauchy surfaces. The first is of topological nature and requires the following fundamental result of Schoen and Yau on the relationship between scalar curvature and topology on a Riemannian $3$-manifold.

\begin{Theorem}\cite[Thm.\ 5.2]{schoen1979existence}
\label{theorem: schoenyautopologyscalarcurvature}
Let $N$ be a compact, oriented $3$-manifold. Suppose that one of the following holds:
\begin{enumerate}
    \item $\pi_1(N)$ contains a finitely generated, non-cyclic Abelian subgroup.
    \item $\pi_1(N)$ contains a subgroup which is isomorphic to the fundamental group of a surface of genus $g > 1$.
\end{enumerate}
Then $N$ admits no Riemannian metric of positive scalar curvature. Moreover, every Riemannian metric on $N$ of nonnegative scalar curvature is flat.
\end{Theorem}

The second proof follows from much simpler Lorentz-geometric arguments and is more generally applicable, see the subsequent Section. It requires the following CMC foliation theorem due to Bartnik.

\begin{Theorem}\cite{bartnik1988remarks}
\label{theorem: CMCfoliation}
Let $(\tilde{M},\tilde{g})$ be a globally hyperbolic spacetime, with compact Cauchy surfaces, satisfying the strong energy condition. Suppose $S \subset \tilde{M}$ is a compact maximal Cauchy surface. Then a neighborhood of $S$ is foliated with CMC Cauchy surfaces. Moreover, if every such foliation is by maximal surfaces, then the spacetime is isometric to an open subset of $(\R \times S, -dt^2 + h)$, where $h$ is the induced Riemannian metric on $S$. In particular, the spacetime is static.
\end{Theorem}

Let us now give the announced proofs:

\begin{Theorem}\cite{bartnik1988remarks}
\label{Theorem: bartniknoCMC}
The constructed spacetime $(M,g)$ is globally hyperbolic, timelike geodesically incomplete and satisfies the strong energy condition. Its Cauchy surfaces are topologically $T^3 \# T^3$ and it does not contain any CMC Cauchy surfaces.
\begin{proof}
The only part left to prove is the nonexistence of CMC Cauchy surfaces.
    \begin{enumerate}
        \item \textit{Topological proof}: Let $\phi:(M,g) \to (M,g)$ be the time inverting isometry. Suppose $S \subset M$ is a compact CMC Cauchy surface. If $H_S \neq 0$, since $H_{\phi(S)} = -H_S$, we may assume w.l.o.g.\ that $H_S > 0$. We claim that $\phi(S) \subset I^+(S)$. Indeed: Suppose $\varphi(S)$ is not a subset of $I^+(S)$. Then there is a point $p \in \phi(S)$ such that $p \notin I^+(S)$. An inextendible future timelike curve through $p$ must intersect $S$, say at $q$. Let $\gamma$ be a maximizing future timelike geodesic from $p$ to $q$. By the Brill-Flaherty uniqueness result (see \cite[Eq.\ (2)]{bartnik1988remarks}), we have
        \[H_{\varphi(S)}(p) - H_S(q) \,\geq\, \int \text{Ric}(\gamma', \gamma').
        \]
        The right-hand side is $\geq 0$, but the left-hand side is $<0$, which is a contradiction. So $\phi(S) \subset I^+(S)$. Now we invoke the maximum principle for prescribed mean curvature \cite{Gerhardt_1983, Bartnik_1984} to conclude the existence of a maximal Cauchy surface $S_0 \subset I^+(S) \cap I^-(\phi(S))$. Since $\pi_1(S_0) \cong \pi_1(T^3 \# T^3) \cong \mathbb{Z}^3 * \mathbb{Z}^3$, it contains the subgroup $\mathbb{Z}^2$ which is non-cyclic, finitely generated and Abelian. Thus, by Theorem \ref{theorem: schoenyautopologyscalarcurvature}, the induced metric $h$ on $S_0$ can only have nonnegative scalar curvature if it is flat. But by the constraint equation for scalar curvature (see \cite[Eq.\ (G.2)]{chrusciel2020geometry}),
        \begin{align*}
            R_h = 16 \pi \rho + |K|_h^2 - H_{S_0}^2,
        \end{align*}
        where $K$ is the second fundamental form. By maximality, $H_{S_0} = 0$, so $R_h \geq 0$. We conclude that $h$ is flat, which implies $R_h = 0$. But $\rho$ is nonzero e.g.,\ in the FLRW parts of the spacetime, which gives the desired contradiction.
        \item \textit{Lorentz-geometric proof}: Suppose $S \subset M$ is a compact CMC Cauchy surface, and suppose for the moment that $H_S \neq 0$. Without loss of generality, we may assume that $H_S > 0$. By Hawking's singularity theorem, each past inextendible timelike curve starting in $S$ has finite Lorentzian arclength. However, this is contradicted e.g.,\ by the ``Lemaître-vertical" curves in Schwarzschild (one could also argue via the infinitely long verticals in the FLRW parts): Consider $\gamma: t \mapsto (t, r_0, p_0)$ (with $r_0 < 1- \varepsilon$, $p_0 \in S^2$ arbitrary) in the $(M^+,g^+)$-part of $(M,g)$: We may extend $\gamma$ so far into the future that it meets $S$, but already the Schwarzschild part has infinite Lorentzian arclength, a contradiction. So the CMC Cauchy surface $S$ must be maximal, $H_S = 0$. By Theorem \ref{theorem: CMCfoliation}, a neighborhood of $S$ is foliated by CMC Cauchy surfaces, all of which must again be maximal. Hence, by the same reference, $(M,g)$ is static, which is manifestly not the case.
    \end{enumerate}
\end{proof}
\end{Theorem}

\begin{Remark}
\label{Remark: RemarktoLorgeomproof}
    Note that the Lorentz-geometric proof does not require the FLRW part to be spatially $T^3$. In fact any compact Riemannian quotient of $\R^3$ does the job. Nor is the application of the ``side-switch" isometry really necessary: Indeed, if $H_S > 0$, we use the infinitely long past curves in the $(M^+,g^+)$-side to arrive at a contradiction to Hawking's singularity theorem, and if $H_S < 0$, in a similar manner we use the infinitely long future curves in the $(M^-,g^-)$-side.
\end{Remark}

\begin{Remark}
In \cite{chrusciel2005initial}, Chru\'sciel, Isenberg and Pollack use their results on the gluing of initial data sets to construct a globally hyperbolic spacetime with $\Ric = 0$ and compact Cauchy surfaces of topology $T^3 \# T^3$, that does not contain any CMC Cauchy surfaces. While Bartnik's example is manifestly timelike geodesically incomplete, this is unclear in their example. However, null geodesic incompleteness has been established, cf.\ \cite{burkhart2019null, burkhart2019causal}. It would be interesting to see whether similar generalizations as the ones we give in the next section could be achieved using initial data gluing methods.
\end{Remark}

\section{Generalizations of Bartnik's ``no-CMC" example}
\label{sec: generalizationsofBartnikexample}

In this Section, we generalize Bartnik's construction by gluing Schwarzschild to hyperbolic FLRW models which have compact spacelike slices. Here, the ODE in Proposition \ref{eq: ODEforY} (with $W > 1$) is not explicitly solvable, but implicit integration will turn out to be sufficient for our purposes. The properties are very similar to the flat case ($W=1$). It turns out that one cannot do this construction with spherical FLRW, because if $W < 1$ the ODE develops finite time past singularities, giving an obstruction to global hyperbolicity.

\begin{Proposition}
\label{Proposition: W > 1 full description}
    Let $(M = D \times S^2, g)$ be a dust-filled Tolman-Bondi spacetime with $\dot{Y} < 0$, monotonically increasing marginal condition $W \geq 1$ and monotonically increasing positive mass function $S > 0$. Then there exists a smooth function $t_0(r)$ such that the metric can be extended smoothly to $\Tilde{D} \times S^2$, where $\Tilde{D} = \{(t,r) \in \R^2 : r \in \{S > 0\}, t \in (-\infty, t_0(r))\}$. Conversely, for any choice of strictly monotonically increasing smooth function $t_0(r)$, monotonically increasing positive functions $S(r)$ and $W(r)$ (with $W(r) \geq 1$), the unique solution of the ODE family $\dot{Y} = - \sqrt{W^2 - 1 + S/Y}$ with initial condition determined by $t_0$ defines a Tolman-Bondi spacetime (together with $X:= Y'/W$) on $\tilde{D} \times S^2$. In particular, $Y' > 0$ on $\Tilde{D}$. Whenever $W(r) = 1$, the solution is given explicitly as described in Proposition \ref{Proposition: ExplicitTB} and whenever $W(r) > 1$, the solution is described implicitly by
    \begin{align*}
        t_0(r) - t = F(W(r),S(r),Y(t,r)).
    \end{align*}
    In this case, $F=F(x,y,z)$ is the function
    \begin{align*}
        F(x,y,z) = \frac{\sqrt{x^2-1 + \frac{y}{z}}z}{x^2 - 1} - \frac{y \, \mathrm{arcoth}\left(\sqrt{1 + \frac{y}{z(x^2 - 1)}} \right)}{(x^2 - 1)^{\frac{3}{2}}}.
    \end{align*}
\begin{proof}
    By Proposition \ref{Proposition: ODEforY}, $Y$ satisfies
    \begin{align*}
        \dot{Y}(t,r) = - \sqrt{W(r)^2 - 1 + \frac{S(r)}{Y(t,r)}}.
    \end{align*}
    This ODE is uniquely solvable upon fixing some initial data $Y(s_0,r)$ for every $r \in \{S > 0\}$. If $W(r) = 1$, we are in the situation of Proposition \ref{Proposition: ExplicitTB}, so we may assume $W > 1$. Let $(a,b) = (a(r),b(r))$ be the maximal existence interval for $t$ when considering $Y$ as a solution of this ODE. Clearly $a = -\infty$ for every $r$, since $Y$ is monotonically decreasing in $t$ and $\dot{Y}$ is bounded for $Y \to + \infty$. To determine $b=b(r)$, we separate variables in the above ODE and integrate:
    \begin{align*}
        t_0(r) - t = \frac{\sqrt{W(r)^2 + \frac{S(r)}{Y(t,r)}} Y(t,r)}{W(r)^2 - 1} - \frac{S(r) \, \mathrm{arcoth}\left(\sqrt{ 1 + \frac{S(r)}{Y(t,r)(W(r)^2-1)}} \right)}{(W(r)^2 - 1)^{\frac{3}{2}}}. \quad (*)
    \end{align*}
    Here, $t_0(r)$ is an integration constant depending on $r$, which can be determined by evaluating $(*)$ at the initial time $t=s_0$. Note that by monotonicity, $Y(t,r) \to 0$ as $t \to b(r)$. Using that $\mathrm{arcoth}(x) \to 0$ as $x \to \infty$, the right hand side of $(*)$ tends to $0$ as $t \to b(r)$. So, necessarily, $b(r) = t_0(r)$.

    It is useful to view the right hand side of $(*)$ as $F(W(r),S(r),Y(t,r))$, where $F=F(x,y,z)$ is a function in three variables defined above and which satisfies, in light of the integration of the ODE,
    \begin{align*}
        \partial_z F(x,y,z) = \frac{1}{\sqrt{x^2 - 1 + \frac{y}{z}}}.
    \end{align*}
    In particular, $\partial_z F(W(r),S(r),Y(t,r)) > 0$. Differentiating $(*)$ with respect to $r$ and solving for $Y'$ yields
    \begin{align*}
        Y'(t,r) = \frac{1}{\partial_z F} \left( t_0'(r) - \partial_x F \cdot W'(r) - \partial_y F \cdot S'\right).
    \end{align*}
    Since $t_0(r) = F(W(r),S(r),Y(s_0,r)) + s_0$, it follows that
    \begin{align*}
        Y'(t,r) = \frac{1}{\partial_z F} ((\partial_x F|_{s_0} - \partial_x F|_t)W' + (\partial_y F|_{s_0} - \partial_y F|_t)S' + \partial_z F|_{s_0} Y'(s_0,r)).
    \end{align*}
    Since $Y'(s_0,r) > 0$ (by positivity of $W$), it follows that $(\partial_z F|_{s_0}) Y'(s_0,r) > 0$. We claim that the other summands are positive as well (recall that $W',S' \geq 0$), as long as $t < s_0$: Indeed,
    \begin{align*}
        \partial_t (\partial_x F|_t) = \partial_x \partial_z F|_t \dot{Y}(t,x),\\
        \partial_t (\partial_y F|_t) = \partial_y \partial_z F|_t \dot{Y}(t,x).
    \end{align*}
    One checks that
    \begin{align*}
        \partial_x \partial_z F(x,y,z) &= -x\left(x^2 - 1 + \frac{y}{z}\right)^{-\frac{3}{2}} < 0,\\
        \partial_y \partial_z F(x,y,z) &= -\frac{1}{2z} \left(x^2 - 1 + \frac{y}{z}\right)^{-\frac{3}{2}} < 0.
    \end{align*}
    Hence together with $\dot{Y} < 0$ we get $\partial_t (\partial_x F(W(r),S(r),Y(t,r))) > 0$ and\\$\partial_t (\partial_y F(W(r),S(r),Y(t,r))) > 0$. Moreover, one can check explicitly that $\partial_x F \to 0$ and $\partial_y F \to 0$ as $z \to 0$, so if the initial condition satisfies
    \begin{align*}
        \partial_z F|_{s_0} Y'(s_0,r) + \partial_x F|_{s_0} W' + \partial_y F|_{s_0} S' > 0,
    \end{align*}
    or equivalently $t_0' > 0$, then $Y'(t,r) > 0$ for all $r$ and for all $t \in (-\infty,t_0(r))$.
\end{proof}
\end{Proposition}

Let us now generalize the construction in the previous section (up to appropriate choices of parameters) arbitrary (compactified) flat or hyperbolic FLRW spacetimes to Schwarzschild. Since for general marginal conditions $W$, dust-filled Tolman-Bondi spacetimes cannot be written as explicitly as in the case $W=1$ (cf.\ Proposition \ref{Proposition: ExplicitTB}), we have to rely on our general gluing result (Theorem \ref{Proposition: GluingTB}) in conjunction with the above analysis on the maximal solution intervals of the ODE in Proposition \ref{Proposition: ODEforY}.

Let us now dive into the details: Let $N$ be either $\R^3$ or $H^3$, let $Q$ be a compact Riemannian quotient manifold of $N$, $q:N \to Q$ the quotient map. Let $x_0 \in Q$ and $0 < 3r_2$ be such that $B^{Q}_{3r_2}(x_0) \subset Q$ is isometric to $B^{N}_{3r_2}(y_0)$ (for any $y_0 \in q^{-1}(x_0)$). Let $(M_2,g_2)$ be the spacetime $(-\infty,r_2) \times_a (Q \setminus B^Q_{r_2}(x_0))$ with $a = a_f$ resp.\ $a = a_h$ if $N = \R^3$ resp.\ $N = H^3$ (we again shift the time parameter for monotonicity reasons). Then the portion $(-\infty,r_2) \times_a (B^Q_{3r_2}(x_0) \setminus B^Q_{r_2}(x_0))$ can be written in Tolman-Bondi form, identical to flat or hyperbolic FLRW. Let $(M_1,g_1)$ be Schwarzschild spacetime in Lemaître coordinates up to a radial coordinate value $r \leq r_1 < r_2$. For fixed Schwarzschild mass $m > 0$, in the flat case the Friedmann mass $\mathcal{M}$ needs to be chosen such that 
\begin{align*}
    S_1(r_1) = 2m \leq \frac{8\pi \mathcal{M}}{3} r_2^3 = S_2(r_2),
\end{align*}
and in the hyperbolic case we need
\begin{align*}
    S_1(r_1) = 2m \leq \frac{8 \pi \mathcal{M}}{3} \sinh(r_2)^3.
\end{align*}
Then it is possible to glue $(M_1,g_1)$ to $(M_2,g_2)$ along the interval $(r_1,r_2)$ by prescribing monotonically increasing connecting functions $W,S$, and a strictly monotonically increasing connecting function $t_0$ for the corresponding functions on either side, cf.\ Proposition \ref{Proposition: W > 1 full description}. We call the resulting spacetime $(M^+,g^+)$.
\begin{Lemma}
\label{Lemma: M+g+ glob hyp in hyp case}
    $(M^+,g^+)$ is globally hyperbolic.
\begin{proof}
    For simplicity, we give the argument for global hyperbolicity if the gluing procedure is done without any quotient operation on the FLRW part in order to use global Tolman-Bondi spacetime arguments, the compactified case can be proven along similar lines.\footnote{Indeed, if the spatial FLRW slice is a quotient $Q$ of $H^3$, consider the universal cover which is partitioned by fundamental domains. Using analogous arguments as the ones appearing in this proof, it can be shown that the future end of the lift to the spacetime universal cover of any inextendible causal curve lies in just one fundamental domain, likewise for the past end. Since the fundamental domains are portions of a Tolman-Bondi spacetime, the arguments given in this proof apply.}
    Keeping this in mind, $(M^+,g^+)$ is a warped product $D \times_{Y} S^2$, so it suffices to check global hyperbolicity of the two-dimensional spacetime $(D,-dt^2 + X^2 dr^2)$, where $D = \{(t,r) \in \R^2 : t < t_0(r)\}$ (cf.\ \cite[Thm.\ 3.68]{BeemEhrlich}). For small enough $\delta > 0$, we claim that $\Sigma:=\{t = t_0(r) - \delta\}$ is a Cauchy surface. It is clearly a smooth, spacelike hypersurface. Let $\gamma(s):=(t(s),r(s))$, $s \in (0,1)$, be an arbitrary $C^1$-causal curve. We will show that if $\gamma$ does not meet $\Sigma$, its limits at $s=0$ or $s=1$ exist. The acausality of $\Sigma$ can be proven along similar lines and will be omitted.

    Suppose first that $t(s) < t_0(r(s)) - \delta$. By monotonicity, $t(s)$ has a limit $T$ as $s \to 1$, with $T \leq r_2 - \delta$ (since $t_0 \leq r_2$ by construction). Again by causality,
    \begin{align*}
        |\dot{r}(s)| \leq \frac{\dot{t}(s)}{X(t(s),r(s))}.
    \end{align*}
    Let $R^+:=\limsup_{s \to 1} r(s)$. Suppose that $R^+ = + \infty$ and let $s_{2,n}$ be a realizing sequence, i.e.,\ $r(s_{2,n}) \to +\infty$. Then the points $(t(s_{2,n}),(r(s_{2,n})))$ are eventually in the hyperbolic FLRW part of $M^+$, where $X(t,r) = a_h(t)$. Let $s_1$ be such that this is true for all $s \in (s_1,s_{2,n})$. Using that $a_h' \geq 1$ by Friedmann's equation, we get
    \begin{align*}
        \dot{r}(s) \leq |\dot{r}(s)| \leq \frac{\dot{t}(s)}{a_h(t(s))} \leq \frac{\dot{t}(s)}{a_h(t(s))} a_h'(t(s)) = \partial_s \log(a_h(t(s))).
    \end{align*}
    Integrating from $s_1$ to $s_{2,n}$, we see that $r(s_{2,n}) \to + \infty$ is impossible. Thus, $R^+ < + \infty$ and due to $T \leq t_0(R^+) - \delta < t_0(R^+)$, $(T,R^+) \in D$. Now let $R^-:=\liminf_{s \to 1} r(s)$. Suppose that $R^- = -\infty$ and let $s_{1,n}$ be a realizing sequence. In this case, $(t(s_{1,n}),r(s_{1,n}))$ is eventually in the Schwarzschild part, so one can argue just as in the flat case to show $R^- > - \infty$. To show that $R^+ = R^-$, observe that $\min_{s \in [s_{1,n},s_{2,n}]}W(r(s))^{-1}$ is bounded below by some constant $C > 0$ independently of $n$. Then, using causality and the fact that $\partial_t Y < 0$,
    \begin{align*}
       (*) \quad  \quad t(s_{2,n}) - t(s_{1,n}) &\geq \int_{s_{1,n}}^{s_{2,n}} \frac{\partial_r Y(t(s),r(s))}{W(r(s))} |\dot{r}(s)| ds\\
        &\geq C (Y(t(s_{2,n},r(s_{2,n}))) - Y(t(s_{1,n},r(s_{1,n})))).
    \end{align*}
    Taking first $s_{2,n} \to 1$ and then $s_{1,n} \to 1$ gives $T - T = 0$ on the left hand side, thus giving
    \begin{align*}
        Y(T,R^-) \geq Y(T,R^+).
    \end{align*}
    Since $\partial_r Y > 0$, we get $R^+ = R^-$. Thus, $\gamma$ is extendible to $s=1$.
    
    The other possibility is that $t(s) > t_0(r(s)) - \delta$ for all $s \in (0,1)$. We will show that $\gamma$ is extendible to $s=0$. Again by monotonicity, $t(s)$ has a limit $T$ as $s \to 0$ (a priori $T \geq - \infty$). Suppose that indeed $T = -\infty$. Let $R^-:=\liminf_{s \to 0} r(s)$. It is impossible that $R^- > -\infty$, since $t_0(r(s)) - \delta < t(s) < t_0(r(s))$. So $R^- = -\infty$. In particular, we may choose a realizing sequence $r(s_n) \to - \infty$ for $s_n \to 0$. Then $\gamma(s_n)$ is eventually in the Schwarzschild region of $(M^+,g^+)$. Thus, causality together with $t(s) > t_0(r(s)) - \delta$ imply the estimate (using the explicit form of the Tolman-Bondi metric for Schwarzschild, cf.\ Example \ref{example: Schwarzschild})
\begin{align*}
    \dot{t}(s_n)^2 \geq \left( \frac{2 m^{1/3}}{(6  \delta)^{1/3}} \right)^2 \dot{r}(s_n)^2.
\end{align*}
Note that everything on the right hand side is bounded independently of $\gamma$. If $\delta$ was chosen small enough, then $t(s_n) \to -\infty$ much faster than $r(s_n) \to -\infty$, thus the inequality $t(s_n) > t_0(r(s_n)) - \delta = r(s_n) - \delta$ cannot be maintained. The conclusion is that $T > -\infty$. From here, one can use $(*)$ and the arguments in the preceding case to see that $+\infty > R^+ := \limsup_{s \to 0} r(s) = R^- =:R$. Moreover, the function $t(s) - t_0(r(s))$ is easily seen to be monotonically increasing in $s$ because $t_0'(r(s)) \leq X(t(s),r(s)) = Y'(t(s),r(s))/W(r(s))$ and $\gamma(s)$ is causal. Hence, $T = t_0(R)$ is impossible. So $T < t_0(R)$, thus $(T,R) \in D$. This shows extendibility of $\gamma(s)$ to $s=0$.
\end{proof}
\end{Lemma}
Similarly, let $\tilde{Q}$ be a compact quotient of either $\R^3$ or $H^3$, and construct the analogous globally hyperbolic spacetime $(M^-,g^-)$, but with inverted time orientation. We extend both to $r = -\infty$ and glue them there to obtain a spacetime $(M,g)$, which is also globally hyperbolic.


\begin{Theorem}[Examples of ``no-CMC" spacetimes]
\label{Theorem: generalizationofBartnikconstruction}
    Let $Q,\Tilde{Q}$ be compact Riemannian quotients\footnote{All combinations are allowed: $Q,\Tilde{Q}$ both quotients of $\R^3$, both quotients of $H^3$, or one a quotient of $\R^3$ and the other a quotient of $H^3$.} of $\R^3$ or $H^3$. Let $(M,g)$ be the spacetime constructed above by gluing Schwarzschild (of any mass $m$) to flat or hyperbolic FLRW which is spatially $Q$ (with Friedmann mass $\mathcal{M}_1 = \mathcal{M}_1(m)$), and then attaching a time-inverted gluing of Schwarzschild to flat or hyperbolic FLRW which is spatially $\Tilde{Q}$ (with Friedmann mass $\mathcal{M}_2 = \mathcal{M}_2(m)$) along the Schwarzschild event horizon. Then $(M,g)$ is globally hyperbolic with topologically $\Tilde{Q} \# Q$ Cauchy surfaces, is timelike geodesically incomplete, satisfies the strong and dominant energy conditions, and does not contain any CMC Cauchy surfaces.
\begin{proof}
    Evidently, $(M^+,g^+)$ and $(M^-,g^-)$ contain infinitely long past and future timelike curves, respectively, so the proof may be carried out just like in the Lorentz-geometric proof of Theorem \ref{Theorem: bartniknoCMC}.

    In the symmetric case of $Q = \Tilde{Q}$ being a quotient of either $\R^3$ or $H^3$ (i.e.,\ $(M^-,g^-)$ is just a time-reversed copy of $(M^+,g^+)$), as well as $Q$ oriented, a proof based on Theorem \ref{theorem: schoenyautopologyscalarcurvature} may also be given. (In the hyperbolic case, we utilize the positive resolution of the surface subgroup conjecture \cite{kahn2012immersing}.) 
    If $Q$ is a quotient of $\R^3$, then $\pi_1(Q)$ contains a subgroup which is isomorphic to the fundamental group of a surface of genus $g=1$. If $Q$ is a quotient of $H^3$, then $\pi_1(Q)$ contains a subgroup which is isomorphic to the fundamental group of a surface of genus $g \geq 2$. In either case, $\pi_1(Q \# Q) \cong \pi_1(Q) * \pi_1(Q)$ contains a subgroup which is isomorphic to the fundamental group of a genus $g \geq 1$ surface. One proceeds as in the topological proof of Theorem \ref{Theorem: bartniknoCMC}, but uses Theorem \ref{theorem: schoenyautopologyscalarcurvature}(i) in the $\R^3$ case and Theorem \ref{theorem: schoenyautopologyscalarcurvature}(ii) in the hyperbolic case.  
\end{proof}
\end{Theorem}

\begin{Remark}[Higher spacetime dimensions]
\label{Remark: higherdimensions}
We have done our analysis of Tolman-Bondi spacetimes and constructed spacetimes without CMC Cauchy surfaces in spacetime dimension $d=4$ in order to be consistent with the literature \cite{bondi1947spherically,eardley1979time,bartnik1988remarks}. However, all of our gluing and ODE arguments continue to hold for $d$-dimensional Tolman-Bondi spacetimes $D \times S^{d-2}$, where the metric is of the form
\begin{align*}
    g = -dt^2 + X(t,r)^2 dr^2 + Y(t,r)^2 d\Omega_{d-2}^2.
\end{align*}
Here $d\Omega_{d-2}^2$ denotes the standard round metric on $S^{d-2}$. Examples of this metric are $\mathrm{FLRW}$ spacetimes in $d$ dimensions, as well as the $d$-dimensional Schwarzschild--Tangherlini spacetime of mass $m > 0$ $(R \neq 0,2m)$
\begin{align*}
    g_S = -\left(1 - \frac{2m}{R^{d-3}}\right) d\tau^2 + \frac{1}{1-\frac{2m}{R^{d-3}}}dR^2 + R^2 d\Omega_{d-2}^2.
\end{align*}
Just like in $d=4$ spacetime dimensions, one can introduce Lemaître coordinates $(t,r)$ via
\begin{align*}
    &dt = d\tau + \sqrt{\frac{2m}{R^{d-3}}} dR^2,\\
    &dr = d\tau + \sqrt{\frac{R^{d-3}}{2m}} dR^2,
\end{align*}
in which the metric takes the form
\begin{align*}
    g_S = -dt^2 + (2m)^{\frac{2}{d-1}} \left(\frac{d-1}{2} (r-t)\right)^{\frac{6-2d}{d-1}} dr^2 + (2m)^{\frac{2}{d-1}} \left(\frac{d-1}{2} (r-t)\right)^{\frac{4}{d-1}} d\Omega_{d-2}^2.
\end{align*}
One can then proceed to glue $d$-dimensional spatially compact flat or hyperbolic FLRW models to Schwarzschild--Tangherlini (where, as before, the Friedmann mass parameter $\mathcal{M}$ has to be chosen accordingly in dependence on the Schwarzschild mass $m$), and attach two such gluings (one of them time-reversed) along the event horizon. The resulting spacetime can be shown to have no CMC Cauchy surfaces via the same Lorentz-geometric arguments as the ones used in the proofs of Theorems \ref{Theorem: bartniknoCMC} and \ref{Theorem: generalizationofBartnikconstruction}. Note that the topological proof is not applicable, as that relies on Theorem \ref{theorem: schoenyautopologyscalarcurvature} which is a result for $3$-manifolds. Let us summarize these observations in following Corollary.
\end{Remark}

\begin{Corollary}[Higher dimensional "no-CMC" examples]
\label{Corollary: higherdimensionalexamples}
Let $d\geq 4$ and let $Q,\tilde{Q}$ be compact Riemannian quotients of $\R^{d-1}$ or $H^{d-1}$. Let $(M,g)$ be the spacetime constructed by gluing $d$-dimensional Schwarzschild--Tangherlini (of mass $m > 0$) to flat or hyperbolic $d$-dimensional FLRW which is spatially $Q$ (with Friedmann mass $\mathcal{M}_1 = \mathcal{M}_1(m)$), and then attaching a time-inverted gluing of $d$-dimensional Schwarzschild--Tangherlini to flat or hyperbolic $d$-dimensional FLRW which is spatially $\Tilde{Q}$ (with Friedmann mass $\mathcal{M}_2 = \mathcal{M}_2(m)$) along the Schwarzschild--Tangherlini event horizon. Then $(M,g)$ is globally hyperbolic, $\dim M = d$, with Cauchy surfaces of topology $\Tilde{Q} \# Q$, is timelike geodesically incomplete, satisfies the strong and dominant energy conditions, and does not contain any CMC Cauchy surfaces.
\end{Corollary}

\begin{Remark}[The spherical case]\label{remark: spherical}
    When gluing Schwarzschild to spherical FLRW, the function $W$ is eventually $< 1$, so $W^2 - 1 < 0$. This causes the solution of the ODE
    \begin{align*}
        \dot{Y}(t,r) = - \sqrt{W(r)^2 - 1 + \frac{S(r)}{Y(t,r)}}
    \end{align*}
    to develop a past singularity in finite time. So the glued spacetime cannot be globally hyperbolic, as one may simply take inextendible curves emerging from Schwarzschild past timelike infinity $i^-$ which cannot cross into the glued region. Of course, one can remove portions of the Schwarzschild region to make it globally hyperbolic, but in this case, the removed portion would be so large that the resulting spacetime could no longer contain timelike curves with infinite length \emph{and} be a neighborhood of the event horizon ($T = -X$ in Kruskal coordinates). Indeed, if both could be achieved, then we could glue the resulting spacetime with a time-inverted copy of itself to produce a globally hyperbolic spacetime with spatial topology $S^3 \# S^3 \cong S^3$ that would contain timelike curves with infinite length. Since the resulting spacetime is spherically symmetric, this contradicts a result of Burnett \cite[Thm.\ 1]{burnett1995lifetimes}.

    On the other hand, one could do the full time-inverted gluing in the spherical case without first removing a portion of the Schwarzschild spacetime to obtain a spacetime $(M,g)$ which is not globally hyperbolic, then restrict attention to the Cauchy development $(\Tilde{M},\tilde{g})$ of some spacelike hypersurface of topology $S^3 \# S^3 \cong S^3$, thus resolving the issue of global hyperbolicity. Of course the methods of proof used so far do not apply to show that $(\Tilde{M},\Tilde{g})$ has no CMC Cauchy surfaces, as they rely on the existence of infinitely long timelike curves, so it would be interesting to determine if $(\Tilde{M}, \Tilde{g})$ contains a CMC Cauchy surface or not.

    Also, let $(M^+,g^+)$ be a toroidal FLRW glued to Schwarzschild, and let $(M^-,g^-)$ be a time-inverted copy of a spherical FLRW glued to Schwarzschild. One can imagine gluing these spacetimes along the event horizon. However, like above, to achieve global hyperbolocity, portions of the manifold would need to be removed. The resulting spacetime has toroidal spatial topology $T^3 \cong T^3 \# S^3$; however, given the remarks above, we cannot conclude that the spacetime has no CMC Cauchy surface since the Lorentz-geometric proof does not carry over: based on the construction, there are either timelike curves that have infinite length to the past or to the future but not both. It would be interesting to find cosmological spacetimes without CMC Cauchy surfaces whose spatial topologies are $T^3$, or prove that none can exist.
    
\end{Remark}

\section{Conclusion \& Outlook}
\label{sec: conclusionoutlook}

In this work, we enlarge the number of 
 known spatial topologies for cosmological spacetimes (i.e., spacetimes with compact Cauchy surfaces and satisfying the strong energy condition) without CMC Cauchy surfaces. Specifically, we show that if $Q$ is any compact Euclidean or hyperbolic 3-manifold and $\Tilde{Q}$ is any other compact Euclidean or hyperbolic 3-manifold, then there are cosmological spacetimes of that type with spatial topologies $Q \# \Tilde{Q}$. (All of these examples are manifestly timelike incomplete,  see Conjecture \ref{Conjecture: CMCBartnik}.)  To obtain our examples, we generalize a gluing construction of Bartnik \cite{bartnik1988remarks}. We glue general Tolman-Bondi spacetimes with variable marginal conditions and use that to construct cosmological spacetimes without CMC Cauchy surfaces. Bartnik's original construction comes from gluing a Schwarzschild Tolman-Bondi spacetime to a flat FLRW spacetime; the gluing procedure in this case is simplified since the marginal condition is the identity $W = 1$ and so the Einstein equations can be integrated, see Proposition \ref{Proposition: ExplicitTB}. For the hyperbolic case, the marginal condition is $W \geq 1$, and the Einstein equations can no longer be integrated; however, maximal solutions can still be constructed implicitly, see Proposition \ref{Proposition: W > 1 full description}.

Our main result, Theorem \ref{Theorem: 
generalizationofBartnikconstruction}, utilizes two different arguments (both due to Bartnik) to establish the nonexistence of a CMC Cauchy surface. The first is a topological argument
using well-known results from Schoen and Yau which forbid certain three-manifolds from having nonnegative scalar curvature. This argument is purely at the initial data level. The second argument is more Lorentz-geometric and uses Hawking's cosmological singularity theorem, but it requires global knowledge of the spacetime. The topological argument only works when $Q \cong \Tilde{Q}$, and we rely on the positive resolution of the surface subgroup conjecture \cite{kahn2012immersing} when $Q$ is a hyperbolic 3-manifold. The Lorentz-geometric argument works in all cases, but it relies on knowledge of the global spacetime. We describe in Remark \ref{Remark: higherdimensions} and Corollary \ref{Corollary: higherdimensionalexamples} how our arguments generalize to produce analogous examples in arbitrary spacetime dimensions. The vacuum cosmological spacetime with spatial topology $T^3 \# T^3$ constructed by  Chru{\'s}ciel, Isenberg, and Pollack in \cite{chrusciel2005initial} makes use of the topological proof alluded here. In light of the positive resolution of the surface subgroup conjecture, it would be interesting if the example in \cite{chrusciel2005initial} generalizes to the more general topologies considered here.

Possible future directions of research include Tolman-Bondi metrics with more general stress-energy tensors (e.g.,\ radiation models), compatibility with results obtained by initial data gluing methods, as well as attempts to find ``no-CMC" cosmological spacetimes where the timelike incompleteness is less obvious (so as to gain more insight into Conjecture \ref{Conjecture: CMCBartnik} resp.\ Conjecture \ref{Conjecture: Bartniksplitting}). Also, models with only axial symmetry could be of interest. Such models may have a chance to produce spacetimes with non CMC Cauchy surfaces provided Burnett's result \cite[Thm.\ 1]{burnett1995lifetimes} does not generalize to the axially symmetric setting. Lastly, as alluded to at the end of Remark \ref{remark: spherical}, our methods cannot be used to establish the nonexistence of CMC Cauchy surfaces in cosmological spacetimes with toroidal spatial topology $T^3$. It would be interesting to find such examples, if any exist.

\section*{Acknowledgments}

Eric Ling is supported by Carlsberg Foundation CF21-0680 and Danmarks Grundforskningsfond CPH-GEOTOP-DNRF151. Argam Ohanyan is funded by the ÖAW-DOC scholarship of the Austrian Academy of Sciences and by the project P33594 of the Austrian Science Fund FWF. He is grateful for the hospitality of the Mathematics Department of the University of Copenhagen, where parts of this work were conducted. The authors appreciate the hospitality of the Fields Institute and the ESI during the ``Thematic Program on Nonsmooth Riemannian and Lorentzian Geometry" and the ``Conference on Non-Regular Spacetime Geometry", respectively. They thank Piotr Chru\'sciel, Gregory Galloway, Michael Kunzinger and Roland Steinbauer for helpful discussions and valuable feedback.

This research was funded in part by the Austrian Science Fund (FWF) [Grant DOI 10.55776/P-33594]. For open access purposes, the authors have applied a CC BY public copyright license to any author accepted manuscript version arising from this submission. 

\addcontentsline{toc}{section}{References}
\bibliographystyle{abbrv}

\end{document}